\documentstyle[11pt]{article}

\def\kon#1#2{\vbox{\halign{##&&##\cr\lower4pt
\hbox{$\scriptscriptstyle\vert$}\hrulefill &\hrulefill\lower4pt
\hbox{$\scriptscriptstyle\vert$}\cr $#1$&$#2$\cr}}}

\def\al{\alpha}

\def\ro{\varrho}

\def\d{\partial}
\def\=d{\,{\buildrel\rm def\over =}\,}

\def\te{\vartheta}
\def\B{\Bigl}

\begin{document}

\title{Inhomogeneous cosmology in the cosmic rest frame without dark stuff }
\author{G\"unter Scharf
\footnote{e-mail: scharf@physik.uzh.ch}
\\ Physics Institute, University of Z\"urich, 
\\ Winterthurerstr. 190 , CH-8057 Z\"urich, Switzerland}

\date{}

\maketitle\vskip 3cm

\begin{abstract} 

We study Einstein's equations with an isotropic but inhomogeneous metric in the cosmic rest frame. The equations are solved perturbatively in the late Universe. The leading plus next-to-leading order results agree with observations without using a cosmological constant or dark matter.


\end{abstract}

\newpage

\section{Introduction}

Various authors have noticed that there is a disturbing asymmetry in our description of nature: gravity is described geometrically, but all other interactions non-geometrically by quantum field theories (see e.g. [1], remark (4.2,5)). It is possible to unify gravity with Maxwell's theory in a 5-dimensional geometric theory (Kaluza-Klein theory), but this did not lead any further. We know today that electromagnetism must be unified with weak interactions in a gauge theory in Minkowski space and not in a unified geometric theory in the sense of Einstein. Then the only way to remove the above exceptional character of gravity is to give up its geometrical interpretation.

At present there is strong support to do so from observations [2]. Taking the absence of dark matter in our Galaxy seriously, a revision of standard general relativity is demanded. As shown in a previous paper [3] the mildest possible revision which solves the dark matter problem is the conservative interpretation of general relativity as a classical field theory in Minkowski space. Here we are following H. Poincar\'e (in ``Science and Hypothesis'') and {\it we consider geometry as a convention}. After 100 years of the geometric dogma it is hard to except this, furthermore there is a high price to pay. It turns out that the nice {\it mathematical} property of diffeomorphism invariance of Einstein's theory cannot be maintained as a {\it physical} principle. This must be discussed in some detail.

Mathematicians are always proud of working without coordinates. But physics is an experimental science and, therefore, coordinates are needed in order to relate quantitative observations to mathematical quantities. It must be stressed that the coordinates are defined by giving a prescription how to measure them. Then it is quite clear that an arbitrary change of coordinates is physically impossible in general. Furthermore, in classical physics the basic laws are always equations between measurable quantities. The corresponding measuring processes are fixed and cannot be changed arbitrarily. One might argue that after defining primary coordinates any new coordinates can be computed mathematically. However a corresponding physical interpretation is generally not possible without additional assumptions.

Here are some examples. The primary coordinates in astrophysics are spherical because one can only measure one distance $r$ (apparent luminosity distance for example) and two angles $\te$ and $\phi$. Of course one then can compute Cartesian coordinates
$$x^1=r\sin\te\cos\phi,\quad x^2=r\sin\te\sin\phi,\quad x^3=r\cos\te.\eqno(1.1)$$
But to interpret these as three rectangular distances, it is necessary to assume Euclidean geometry. The definition of time is another problem. The signals of atomic clocks depend on their motion and on the local gravitational field, so they are not very useful in the large. The simplest way is to reduce time to length by giving the light speed a prescribed value $c=1$. When astronomers use lightyears they just do this. A severe problem arises when cosmologists use comoving coordinates. Then detailed assumptions about the motion of observers or massive test bodies must be made. On the other hand these motions are used to define the gravitational field (the Christoffel symbols in the geodesic equations). Then the definition of comoving coordinates is logically problematic because there is no operational definition. A second drawback is the local nature of the comoving coordinates; one has to use coordinate patches to cover the whole universe. We shall only use global spherical coordinates in a fixed operationally defined system and we avoid
transformation to other coordinates if there is no clear physical interpretation. The diffeomorphism invariance is not lost, but it ``only'' means that all basic equations are tensor equations.

The preferred frame of reference for the Universe is the cosmic rest frame which is at rest with respect to the cosmic microwave background. Theorists do not like this frame because then all galaxies are in radial motion instead of being at rest in the comoving coordinates. As a consequence the solution of Einstein's equations is much more complicated. Indeed we have found in a previous paper [4], that the metric tensor in the cosmic rest frame has off-diagonal components
$$ds^2=dt^2+2b(t,r)dt\,dr-a^2(t,r)[dr^2+r^2(d\te^2+\sin^2\te d\phi^2)].\eqno(1.2)$$
Furthermore we have noticed in [4] that there remains a dependence on the radial coordinate $r$ in Einstein's equations. For this reason we have to allow a $r$-dependence in the metric functions $a$ and $b$, so that the Universe is no longer homogeneous. Another reason for an inhomogeneous metric is the following. When a big bang has occurred at $t=0$, $r=0$ in the cosmic rest frame, then points with $r>0$ remain still in silence for some time. Consequently the metric is certainly inhomogeneous for early times. In the standard description by comoving coordinates and a homogeneous metric this fact is not visible. Various authors have investigated cosmological inhomogeneities [5] [8], In the recent book by Ellis, Maartens and McCallum [9] numerous models for inhomogeneous cosmology are discussed. But the solution we are going to construct is not contained. Furthermore, in the cosmic rest frame the Universe looks totally different.

The other motivation for this study is the increasing direct evidence that dark matter does not exist in our Galaxy [2]. There may exist a large population of faint stars which doubles the density of visible matter [10]. Still the total mass density remains well below the critical density. Therefore the widely excepted $\Lambda$CDM cosmology is in difficulty. For this reason it is worthwhile to reconsider the expansion in the Universe by starting from scratch.

The paper is organized as follows. In the next section we set up Einstein's equations for the inhomogeneous metric (1.2). In section 3 we consider the homogeneous special case where the $r$-dependence is neglected. We show that this leads to a contradiction in the energy-momentum tensor. To avoid this contradiction we are led in Sect.4 to a perturbative treatment of the full inhomogeneous equations. The results in leading order are similar to the Einstein - de Sitter universe, however interpreted in the cosmic rest frame. It is well known that this solution gives a too big matter density. In the standard $\Lambda$CDM model this is cured by introducing a cosmological constant $\Lambda$ and dark matter. Instead of this we simply go to the next order in Sect.5 and 6. We obtain the  energy-momentum tensor of the inhomogeneous universe and a new integration constant allows to fit the observed normal matter density exactly. In the last section we discuss the application of the theory to the present universe. When the reader has reached this point he perhaps shares the view of the author that the $\Lambda$CDM model is too simple to describe our Universe.

\section{Einstein's equations}

We choose spherical coordinates $t,r,\te,\phi$ which are assumed to be dimensionless, that means the measured values have been divided by suitable units. The components of the metric tensor corresponding to the line element (1.2) are
$$g_{00}=1,\quad g_{01}=b(t,r),\quad g_{11}=-a^2(t,r)\eqno(2.1)$$
$$g_{22}=-r^2a^2(t,r),\quad g_{33}=-r^2a^2(t,r)\sin^2\te,\eqno(2.2)$$
and zero otherwise. The components of the inverse metric are equal to
$$g^{00}={a^2\over D},\quad g^{01}={b\over D},\quad g^{11}=-{1\over D}\eqno(2.3)$$
$$g^{22}=-{1\over a^2r^2},\quad g^{33}=-{1\over a^2r^2\sin^2\te}\eqno(2.4)$$
where
$$D=a^2+b^2\eqno(2.5)$$
is the determinant of the $2\times 2$ matrix of the $t$, $r$ components. The non-vanishing Christoffel symbols are given by
$$\Gamma^0_{00}={b\dot b\over D},\quad \Gamma^0_{01}=-{ab\over D}\dot a,\quad
\Gamma^0_{11}={1\over D}(a^3\dot a-aba'+a^2b')\eqno(2.6)$$
$$\Gamma^0_{22}={r\over D}(ra^3\dot a+ba^2+rbaa'),\quad\Gamma^0_{33}={r\sin^2\te\over D}(ra^3\dot a+ba^2+rbaa')
$$
$$\Gamma^1_{00}=-{\dot b\over D},\quad\Gamma^1_{01}={a\dot a\over D},\quad \Gamma^1_{11}={1\over D}(ab\dot a+aa'+bb')$$
$$\Gamma^1_{22}={r\over D}(rab\dot a-a^2-raa'),\quad\Gamma^1_{33}={r\over D}(rab\dot a-a^2-raa')\sin^2\te,\eqno(2.7)$$
$$\Gamma^2_{02}={\dot a\over a},\quad\Gamma^2_{12}={a'\over a}+{1\over r},\quad\Gamma^2_{33}=-\sin\te\cos\te\eqno(2.8)$$
$$\Gamma^3_{03}={\dot a\over a},\quad\Gamma^3_{13}={a'\over a}+{1\over r},\quad\Gamma^3_{23}={\cos\te\over\sin\te}.\eqno(2.9)$$
Here the dot means $\d/\d t$ and the prime $\d/\d r$.

To write down Einstein's equations we have to calculate the Ricci tensor
$$R_{\mu\nu}=\d_\al\Gamma^\al_{\mu\nu}-\d_\nu\Gamma^\al_{\mu\al}+\Gamma^\beta_{\mu\nu}\Gamma^\al_{\al\beta}
-\Gamma^\al_{\nu\beta}\Gamma^\beta_{\al\mu}$$
using the Christoffel symbols (2.6-9). We obtain
$$R_{00}=-{\ddot a\over D}\B(3a+2{b^2\over a}\B)-{b^2\dot a^2\over D^2}+{\dot a\dot b\over D^2}\B(3ab+2{b^3\over a}\B)
+{\dot b\over D^2}(aa'+bb')+$$
$$-{\dot b'\over D}-2{a'\dot b\over aD}-{2\over r}{\dot b\over D}\eqno(2.10)$$
$$R_{01}=-\ddot a{ab\over D}-\dot a^2{b\over D^2}(2a^2+3b^2)+{ab^2\over D^2}\dot a\dot b-{2b^2\dot a\over raD}+{b^2\over D^2}\dot bb'-{b\over D}\dot b'+$$
$$+{ab\over D^2}a'\dot b+2{\dot a\over D}a'-2{\dot a'\over a}.\eqno(2.11)$$
$$R_{11}={a^3\over D}\ddot a+{a^4\dot a^2\over D^2}\B(2+3{b^2\over a^2}\B)-{a^3b\over D^2}\dot a\dot b
+{2\over rD}(ab\dot a+aa'+bb')+{a^2\over D}\dot b'+$$
$$+2{a\over D}\dot ab'-{a^2b\over D^2}\dot bb'-2{a''\over a}+{2\over D}a'^2+2{b\over aD}a'b'-{4\over r}{a'\over a}\eqno(2.12)$$
$$R_{22}=r^2\B[{a^3\over D}\ddot a+{a^4\dot a^2\over D^2}\B(2+3{b^2\over a^2}\B)-{a^3b\over D^2}\dot a\dot b\B]
+{b^2\over D}+r{ab\over D^2}\dot a(3a^2+4b^2)+$$
$$+r{a^4\over D^2}\dot b+{r^2a^2\over D^2}b'\B(a\dot a+{b\over r}\B)+{r^2\over D}\B(2b\dot aa'+2ba\dot a'
-{4\over r}aa'+a'^2-aa''\B)+$$
$$+{r^2\over D^2}\B(a^3a'\dot b-2a^2b\dot aa'+a^2a'^2+{a^3\over r}a'+aba'b'\B)\eqno(2.13)$$
$$R_{33}=R_{22}\sin^2\te.\eqno(2.14)$$
This gives the following scalar curvature
$$R=g^{\mu\nu}R_{\mu\nu}=-6{a\ddot a\over D}-{\dot a^2\over D^2}(6a^2+12b^2)+$$
$$+6\dot a\dot b{ab\over D^2}-{4\over rD^2}\B(a^2\dot b+2ab\dot a+3{b^3\over a}\dot a\B)
-2{b^2\over r^2a^2D}+2{b\over D^2}\dot bb'-2{\dot b'\over D}-$$
$$-4{b'\dot a\over D^2}a-4{b\over rD^2}b'+{a'\dot b\over D^2}({ab^2\over D}-3a)-8{b\over aD}\dot a'+4{b\over D^2}\dot aa'(3+{b^2\over a^2})+4{a''\over aD}-$$
$$-{a'^2\over D^2}(6+2{b^2\over a^2})-4{b\over aD^2}a'b'+4{a'\over rD^2}(2a+3{b^2\over a}).\eqno(2.15)$$

Next we calculate the Einstein tensor:
$$G_{00}=R_{00}-{g_{00}\over 2}R={\dot a^2\over D^2}\B(3a^2+5b^2\B)-2{\ddot ab^2\over aD}+2{\dot a\dot b\over D^2}{b^3\over a}-$$
$$-2{\dot bb^2\over rD^2}+2{\dot ab\over rD^2}(2a+3{b^2\over a})+2{a\dot a\over D^2}b'
-2{a'b\over D^2}{b^2\over a}+4{b\over aD}\dot a'+{2b\over D^2}\dot aa'\B({b^2\over a^2}-1\B)+$$
$$+2{a''\over aD}+{a'^2\over D^2}\B(1-{b^2\over a^2}\B)
+2{b\over aD^2}a'b'-{2a'\over rD^2}\B(2a+3{b^2\over a}\B)+{b^2\over r^2a^2D}+2{bb'\over rD^2}.\eqno(2.16)$$
$$G_{01}=R_{01}-{g_{01}\over 2}R=b\B[2\ddot a{a\over D}+{\dot a^2\over D^2}(a^2+3b^2)-2\dot a\dot b{ab\over D^2}\B]+$$
$$+{2b^2\over raD^2}\dot a(a^2+2b^2)+2{a^2b\over rD^2}\dot b+2{ab\over D^2}\dot ab'+$$
$$+2{a'\dot b\over D^2}ab+{2\dot a'\over aD}(b^2-a^2)+2{\dot aa'\over D^2}\B(
{b^4\over a^2}-b^2\B)-2{b\over aD}a''+$$
$$+{b\over D^2}a'^2\B(1-{b^2\over a^2}\B)+{2b^2\over aD^2}a'b'-2{a'b\over rD^2}\B(2a+3{b^2\over a}\B)
+2{b^2\over rD^2}b'+{b^3\over r^2a^2D}.\eqno(2.17)$$
$$G_{11}=a^2\B[-2{\ddot aa\over D}-{\dot a^2\over D^2}\B(a^2+3b^2\B)+2\dot a\dot b{ab\over D^2}\B]-{\dot a\over rD^2}(2a^3b+4ab^3)-$$
$$-{2a^4\over rD^2}\dot b+2{ab^2\over D^2}\dot ab'-{b^2\over r^2D}+2{b^3\over rD^2}b'+{2\over a}a''
+{a'^2\over D^2}(3b^2+a^2)+$$
$$+2{b^3\over aD^2}a'b'+{2a'\over raD^2}(a^4-2b^4)-2{a'a^3\dot b\over D^2}-4{ab\over D}\dot a'
+2{b\over D^2}\dot aa'(a^2-b^2).\eqno(2.18)$$
$$G_{22}=r^2a^2\B[-2{a\over D}\ddot a-{\dot a^2\over D^2}\B(a^2+3b^2\B)+2\dot a\dot b{ab\over D^2}\B]
+r^2\B[-\dot a{ab\over rD^2}(a^2+2b^2)-$$
$$-{a^4\over rD^2}\dot b-{a^2\over D^2}b'(a\dot a+{b\over r})+{a^2b\over D^2}\dot bb'-{a^2\over D}\dot b'
+2{\dot aa'\over D^2}ba^2-2{ab\over D}\dot a'+$$
$$+{aa'\over rD^2}(a^2+2b^2)-{a^2\over D^2}a'^2+{a\over D}a''-{ab\over D^2}a'b'\B].\eqno(2.19)$$
We notice that the leading terms in $G_{01}, G_{11}$ and $G_{22}$ agree up to some factors. This will play an important role in the following. Our aim now is to solve Einstein's equations
$$G_{\mu\nu}=8\pi GT_{\mu\nu}.\eqno(2.20)$$

\section{Failure of homogeneous cosmology}

One might think that for large distance from the point $r=0$ where the big bang has occurred the $r$-dependence disappears.
Therefore we consider the special case of a homogeneous cosmology putting $a'=0=b'$ and $r\to\infty$. Then we get from (2.17)
$$G_{01}=b\B(2\ddot a{a\over D}+{\dot a^2\over D^2}(a^2+3b^2)-2{ab\over D^2}\dot a\dot b\B)\eqno(3.1)$$
and (2.19) gives essentially the same result after division by $r^2$:
$${G_{22}\over a^2r^2}=-2\ddot a{a\over D}-{\dot a^2\over D^2}(a^2+3b^2)+2{ab\over D^2}\dot a\dot b.\eqno(3.2)$$
By (2.20) this implies a strange relation between two components of the energy-momentum tensor
$$T_{01}=-{b\over ar^2}T_{22}.\eqno(3.3)$$

In $T_{\mu\nu}$ we only consider ordinary matter and radiation. We first set up the radiation tensor $T^r_{\mu\nu}$.
 Since the metric (1.2) has off-diagonal elements the energy-momentum tensor of radiation in the corresponding gravitational field must have off-diagonal elements as well. Therefore we assume the mixed tensor to be of the form
$$T^{r0}_0=\ro_r,\quad T^{r0}_1=q=T^{r1}_0,\quad T^{r1}_1=T^{r2}_2=T^{r3}_3=-p_r.\eqno(3.4)$$
and zero otherwise. The value of $q$ follows from the basic property that $T^r_{\mu\nu}=T^r_{\nu\mu}$ must be symmetric: since
$$T^r_{01}=g_{01}T^{r1}_1+g_{00}T^{r0}_1=-bp_r+q\eqno(3.5)$$
$$T^r_{10}=g_{10}T^{r0}_0+g_{11}T^{r1}_1=b\ro_r-a^2q\eqno(3.6),$$
we get
$$q={b\over a^2+1}(\ro_r+p_r).\eqno(3.7)$$
The remaining components are
$$T^r_{00}=g_{00}T^{r0}_0+g_{01}T^{r1}_0=\ro_r+bq$$
$$T^r_{11}=a^2p_r+bq\eqno(3.8)$$
$$T^r_{22}=r^2a^2p_r,\quad T^r_{33}=r^2a^2p_r\sin^2\te.$$

In our model of the universe the ordinary matter with proper energy density $\ro_m$ and pressure $p_m$ is in radial motion with velocity $v_m$. In previous attempts we have assumed that $v_m$ can be calculated by solving the geodesic equations. This was not correct for the following reason. The geodesic equation describes the motion of a test body which does not disturb the gravitational field. On the other hand the moving normal matter with density $\ro_m$, pressure $p_m$ and 4-velocity
$$u^\mu=(u^0, u^0v_m,0,0).\eqno(3.9)$$
strongly influences the gravitational field. So its motion is governed by hydrodynamic equations in the gravitational field following from the conservation of the energy-momentum tensor [7]
$$T_m^{\mu\nu}=-p_mg^{\mu\nu}+(p_m+\ro_m)u^\mu u^\nu.\eqno(3.10)$$
which reads
$$\nabla_\mu T_m^{\nu\mu}={\d T_m^{\nu\mu}\over\d x^\mu}+\Gamma^\nu_{\mu\lambda}T_m^{\lambda\mu}+\Gamma^\mu_{\mu\lambda}T_m^{\nu\lambda}=0.\eqno(3.11)$$

It is well known [7] that due to the Bianchi identities Einstein's equations imply the conservation of the total energy-momentum tensor
$$\nabla_\mu(T^{\nu\mu}_r+T^{\nu\mu}_m)=0.\eqno(3.12)$$
So if we can neglect radiation or explicitly guarantee
$$\nabla_\mu T_r^{\nu\mu}=0\eqno(3.13)$$
then the hydrodynamic equations are contained in Einstein's equations, therefore, they need not be imposed separately. This leads to a strategy which we will use in the following: We use Einstein's equations not only to determine $a(t,r)$ and $b(t,r)$ but also to find the most important part of the energy-momentum tensor. The latter is a simple task because no partial differential equation must be solved.

Let us now return to the strange relation (3.3). We consider the late Universe where the matter is non-relativistic so that we put $p_m=0$. Then the matter tensor becomes
$$T_m^{\mu\nu}=\ro_m u^\mu u^\nu.\eqno(3.14)$$
Here $u^0$ follows from the normalization
$$g_{\mu\nu}u^\mu u^\nu=(u^0)^2+2b(u^0)^2v_m-a^2(u^0)^2v_m^2=1.\eqno(3.15)$$
We obtain
$$u^0=(1+2bv_m-a^2v_m^2)^{-1/2}.\eqno(3.16)$$
We also need the covariant components
$$u_0=g_{00}u^0+g_{01}u^1=(1+bv_m)(1+2bv_m-a^2v_m^2)^{-1/2}\eqno(3.17)$$
$$u_1=g_{10}u^0+g_{11}u^1=(b-a^2v_m)(1+2bv_m-a^2v_m^2)^{-1/2}.\eqno(3.18)$$
This gives
$$u_0u_1=(1+bv_m)(b-a^2v_m)(1+2bv_m-a^2v_m^2)^{-1}\equiv wb\eqno(3.19)$$
where
$$w={1+(b-a^2/b)v_m-a^2v_m^2\over 1+2bv_m-a^2v_m^2}$$
is of the order 1. Now the strange relation (3.3) becomes
$$w\ro_m=-p_r=-{\ro_r\over 3}\eqno(3.20)$$
where we have assumed the normal equation of state for the radiation.
In the present Universe with $\ro_m\gg \ro_r$ this relation is clearly violated. The only way to escape the strange relation (3.3) is that the expression (3.2) which appears also in (3.1) vanishes in leading order. This then leads to the inhomogeneous universe which is considered in the following.

\section{An inhomogeneous universe} 

Because of the very many terms in the Einstein tensor it seems to be rather hopeless to solve Einstein's equations exactly. But a perturbative treatment is possible. In the late Universe which we should understand first, $a(t,r)$ and $b(t,r)$ are big ($\gg 1$). Then we put
$$a(t,r)=a_0+a_1+\ldots$$
$$b(t,r)=b_0+b_1+\ldots\eqno(4.1)$$
where $a_0, b_0$ satisfy
$$-2{a_0^3\over D_0}\ddot a_0-\dot a_0^2{a_0^2\over D_0^2}(a_0^2+3b_0^2)+2{a_0^3b_0\over D_0^2}\dot a_0\dot b_0=0\eqno(4.2)$$
with $D_0=a_0^2+b_0^2$. Then the leading terms in $G_{01}, G_{11}$ and $G_{22}$ vanish. Multiplying by $D_0^2/(a_0^3\dot a_0)$ we get a linear equation for $y=b_0^2$:
$$\dot y-y\B(2{\ddot a_0\over\dot a_0}+3{\dot a_0\over a_0}\B)-2{\ddot a_0\over\dot a_0}a_0^2-a_0\dot a_0=0.\eqno(4.3)$$

A special solution is $y_1=-a_0^2$. Since the solution of the homogeneous equation is $C\dot a_0^2a_0^3$, the general solution of (4.3) is equal to
$$y=C\dot a_0^2a_0^3-a_0^2.\eqno(4.4)$$
This must be positive for arbitrary $t,r$. The simplest way to chieve this is to choose $a_0$ independent of $r$ and satisfying
$$\dot a_0^2a_0^3=\lambda a_0^2\eqno(4.5)$$
with $\lambda>1/C$ and constant.The solution of this equation is
$$a_0(t)=\al t^{2/3}\eqno(4.6)$$
where $\al$ is a new constant of integration. This choice of $a_0$ defines our perturbative scheme, $\al$ is the perturbative parameter in (4.1). It is interesting to note that the $t^{2/3}$-law is the same as in the matter dominated universe in the standard FRW cosmology [6], but we have obtained this law from the vacuum equations; furthermore this law gets modified in the next order. We shall see below that the energy-momentum tensor does not contribute to the leading order equation (4.2). From (4.4) we now find the following form of $b_0(t,r)$
$$b_0(t,r)=f(r)a_0(t).\eqno(4.7)$$

To determine $f(r)$ we consider
$$G_{11}-{G_{22}\over r^2}={a^2\over D}\dot b'+{\dot ab'\over D^2}(2ab^2+a^3)-{\dot bb'\over D^2}a^2b$$
$$-\dot b{a^4\over rD^2}-{\dot a\over rD^2}(a^3b+2ab^3)-2{ab\over D}\dot a'-2{\dot aa'\over D^2}b^3+$$
$$+{b'b\over rD^2}(2b^2+a^2)-{b^2\over r^2D}+{a''\over aD}(a^2+2b^2)+{a'^2\over D^2}(2a^2+3b^2)+$$
$$+{a'b'\over D^2}\B(2{b^3\over a}+ab\B)+{a'\over rD^2}\B(a^3-4{b^4\over a}-2ab^2\B)-{a'\dot b\over D^2}a^3=O(\al).\eqno(4.8)$$
After multiplying with $D^2/a^2$ we restrict to the leading order $O(\al^3)$:
$$\dot b'_0D_0+\dot a_0b'_0\B(2{b_0^2\over a_o}+a_0\B)-b'_0\dot b_0b_0-{1\over r}\B[\dot b_0a_0^2+\dot a_0\B(a_0b_0+2{b_0^3
\over a_0}\B)\B]=0.\eqno(4.9)$$
We have used the fact that $a'_0=0$ and again, as well shall see, there is no contribution from $T_{\mu\nu}$. Inserting (4.7) we get a simple equation for $f(r)$
$$f'(r)-{1\over r}f(r)=0$$
with the solution $f(r)=Lr$. Hence
$$b_0(t,r)=Lra_0(t)=Lr\al t^{2/3}\eqno(4.11)$$
is the leading order of $b(t,r)$ (4.1) with another constant of integration $L$ of dimension of a reciprocal length.

The Einstein's equations for $G_{01},G_{11}$ and $G_{22}$ are now fulfilled in leading order. There remains $G_{00}$ to be investigated. Since $G_{00}=O(\al^0)$ according to (2.16) here $T_{00}$ does contribute. 
Using the leading order results we obtain
$$8\pi GT_{00}={\dot a_0^2\over D_0^2}(3a_0^2+5b_0^2)-2{\ddot a_0b_0^2\over a_0D_0}+2{\dot a_0\dot b_0\over D_0^2}
{b_0^2\over a_0}=$$
$$={\dot a_0^2\over D_0^2}a_0^2(3+5L^2r^2+2L^4r^4)-2{\ddot a_0\over D_0}a_0L^2r^2.\eqno(4.12)$$
Substituting (4.6) we get the very simple result
$$T_{00}={1\over 6\pi G t^2},\eqno(4.13)$$
that means in lowest order the inhomogeneity drops out here. This result for $T_{00}$ is the same as in the Einstein - de Sitter universe. However, our energy density $T_{00}$ contains a 4-velocity $u_0>1$ so that (4.13) differs considerably from the matter density $\ro_m$. It is well known that taking for $t$ the present age $T$ of the Universe, the matter density in the Einstein - de Sitter model comes out much too big. To remove the defect one usually introduces a cosmological constant $\Lambda$ and in addition some dark matter [6]. We shall not do so but simply go to :

\section{Next to leading order}

This is also called first order correction to the leading zeroth order because it involves $a_1$ and $b_1$ in (4.1). Now we have to be careful about the contributions of $T_{\mu\nu}$. To get an idea of the radiation contribution we investigate energy conservation
$$\nabla_\mu T_r^{0\mu}={\d T_r^{0\mu}\over\d x^\mu}+\Gamma^0_{\mu\nu}T_r^{\nu\mu}+\Gamma^\mu_{\mu\nu}T_r^{0\nu}=0.\eqno(5.1)$$
We cannot set up energy conservation for the matter content because we do not know the radial velocity $u^1=u^0v_m$. As discussed above (3.12) this is no harm. Using the $\Gamma$'s of Sect.1 we have
$$\d_tT_r^{00}+\d_rT_r^{01}+{b\dot b\over D}T_r^{00}-2{ab\over D}\dot aT_r^{01}+{1\over D}(a^3\dot a-aba'+a^2b')T_r^{11}+$$
$$+{r\over D}(ra^3\dot a+ba^2+raba')T_r^{22}+{r\over D}\sin^2\te(ra^3\dot a+ba^2+raba')T_r^{33}+$$
$$+\B({b\dot b\over D}+{a\dot a\over D}+2{\dot a\over a}\B)T_r^{00}+\B({aa'+bb'\over D}+2{a'\over a}+{2\over r}\B)T_r^{01}=0.
\eqno(5.2)$$

Now we substitute the leading order results. From (3.8) we get
$$T_r^{00}={\ro_r\over 1+L^2r^2}+O(\al^{-2})$$
$$T_r^{01}=Lr{\ro_r\over a(1+L^2r^2)}+O(\al^{-3})\eqno(5.3)$$
$$T_r^{11}={p_r\over a^2(1+L^2r^2)}+{L^2r^2\over a^2+1}{\ro_r+p_r\over 1+L^2r^2}.$$
Then in (5.2) the factor $1/(1+L^2r^2)$ cancels and we arrive at
$$\d_t\ro_r+3{\dot a\over a}(p_r+\ro_r)+O(\al^{-1})=0.\eqno(5.4)$$
This is the same equation as in standard FRW-cosmology [6]. With the usual equation of state $p_r=\ro_r/3$ one has
$$\ro_r(t)={\ro_0\over a^4(t)}\eqno(5.5)$$
where $\ro_0$ is constant.

We start the first order calculation with
$${G_{22}\over a^2r^2}=-{1\over D}(2a\ddot a+\dot a^2)+{2\over D^2}(ab\dot a\dot b-b^2\dot a^2)-$$
$$-{\dot ab\over raD^2}(a^2+2b^2)-\dot b{a^2\over rD^2}-{b'\over D^2}(a\dot a+{b\over r})+\dot bb'{b\over D^2}-{\dot b'\over D}+$$
$$+2{\dot aa'\over D^2}b-2{b\over aD}\dot a'+O(\al^{-2}).\eqno(5.6)$$
We insert (4.1) and collect the first order contributions $O(\al^{-1})$:
$$\B({G_{22}\over a^2r^2}\B)_1=-{1\over D_0}(2a_1\ddot a_0+2a_0\ddot a_1+2\dot a_0\dot a_1)-{2\over D_0^2}(2b_0b_1\dot a_0^2
+2b_0^2\dot a_0\dot a_1)+$$
$$+{8\over D_0^2a_0\beta}(a_1+Lrb_1)(b_0^2\dot a_0^2-\dot a_0\dot b_0a_0b_0)+{2\over D_0^2}(\dot a_1\dot b_0a_0b_0+\dot b_1
\dot a_0a_0b_0+$$
$$+a_1\dot a_0\dot b_0b_0+b_1\dot a_0\dot b_0a_0)
-{\dot a_0b_0\over ra_0D_o^2}(a_0^2+2b_0^2)-\dot b_0{a_0^2\over rD_0^2}-L{a_0^2\over D_0^2}\dot a_0+$$
$$+\dot b_0La_0{b_0\over D_0^2}-L{\dot a_0\over D_0}+O(\al^{-2})=8\pi Gp_r.\eqno(5.7)$$
We assume pressureless matter so that only radiation pressure contributes on the right-hand side. In (5.7) we have used the expansion
$$D=D_0\B(1+{2\over a_0}{a_1+Lrb_1\over \beta}\B)\eqno(5.8)$$
and the abbreviation
$$\beta=1+L^2r^2.\eqno(5.9)$$
The right-hand side in (5.7) must be of order $\al^{-1}$. Therefore we write (5.5) in the form
$$p_r={\ro_r\over 3}={p_0\over a_0t^2}+O(\al^{-2}),\eqno(5.10)$$
where $p_0$ is a new constant of integration which contains $\al^2$: $p_0=\ro_0/3\al^3$.

Substituting the lowest order results into (5.7) we arrive at
$$\ddot a_1+{\dot a_1\over t}\B({4\over 3}-{2\over 3\beta}\B)-{a_1\over t^2}\B({2\over 3}-{4\over 9\beta}\B)-{\dot b_1\over t}{2\over 3} {Lr\over\beta}+$$
$$+{b_1\over t^2}{4Lr\over 9\beta}+{2L\over 3t}\B({1\over\beta}+1\B)+8\pi G{\beta p_0\over 2t^2}=0.\eqno(5.11)$$
To solve the time-dependence we set
$$a_1(t,r)=tg_1(r)+h_1(r)\eqno(5.12)$$
$$b_1(t,r)=tg_2(r)+h_2(r).\eqno(5.13)$$
Now the terms proportional to $t^{-1}$ give
$$-(3L^2r^2+2)g_1+Lrg_2=3L(\beta+1)$$
or
$$g_2=(3Lr+{2\over Lr})g_1+{3\over r}(L^2r^2+2)\equiv \gamma_1g_1+\gamma_2.\eqno(5.14)$$
The terms proportional to $t^{-2}$ lead to
$$h_1\B({4\over 9\beta}-{2\over 3}\B)+h_2{4Lr\over 9\beta}+4\pi G\beta p_0=0$$
or
$$h_1={2\over 3L^2r^2+1}(Lrh_2+9\pi G\beta^2p_0).\eqno(5.15)$$

Next we consider
$${1\over b}G_{01}+{G_{22}\over a^2r^2}={\dot ab\over rD^2}\B(a+2{b^2\over a}\B)+\dot b{a^2\over rD^2}+b'{a\dot a\over D^2}+$$
$$+2{a\dot b\over D^2}a'-2{\dot a'\over D}\B({a\over b}+{b\over a}\B)-{\dot b'\over D}+{b\over D^2}\dot bb'+$$
$$+2{\dot aa'\over D^2}{b^3\over a^2}-{a''\over aD}-{a'^2\over D^2}{b^2\over a^2}+{b\over D^2}{a'\over a}b'-$$
$$-{a'\over rD^2}\B(3a+4{b^2\over a}\B)+{bb'\over rD^2}+{b^2\over ra^2D}+O(\al^{-3})=8\pi G{T^m_{01}\over b}.\eqno(5.16)$$
This combination has been chosen in such a way that radiation does not contribute. In lowest order this gives
$${2b_0\over ra_0D_0^2}\dot a_0(a_0^2+2b_0^2)+2\dot b_0{a_0^2\over rD_0^2}+2{a_0\over D_0^2}\dot a_0b_0'
=8\pi G{T^m_{01}\over b}.\eqno(5.17)$$
Using (4.6) and (4.7) we finally obtain
$$8\pi GT^m_{01}={4\over 3}{L^2r\over\beta t}\eqno(5.18)$$
which is of the order $O(\al^0)$. 

We treat $G_{11}-G_{22}/r$ (4.8) in the same way, here we need $T^m_{11}$. By (3.14) this can now be calculated according to
$$T_{11}^m={(T^m_{01})^2\over T^m_{00}}\eqno(5.19)$$
which yields
$$8\pi GT^m_{11}={4\over 3}{L^4r^2\over\beta^2}.\eqno(5.20)$$
Substituting the leading order expressions we finally obtain
$$\B(G_{11}-{G_{22}\over r^2}\B)_1={1\over\beta}\dot b_1'-{2\over\beta}Lr\dot a_1'+{2\over 3t\beta}b_1'-{4L^3r^3\over 3t\beta^2}a_1'
-{1\over\beta r}\dot b_1-$$
$$-{2\over 3\beta tr}b_1+{L^4r^2\over\beta^2}=8\pi GT_{11}^m.\eqno(5.21)$$
Inserting (5.20), substituting (5.12-13) and separating the $t$-dependence we obtain the following ODE for the $g$'s
$${5\over 3}g_2'-{5\over 3r}g_2-g_1'\B(2Lr+{4L^3r^3\over 3\beta}\B)=
{L^2\over 3}\B(1-{1\over\beta}\B),\eqno(5.22)$$
and in addition we get a homogeneous equation for the $h$'s:
$$h_2'-{h_2\over r}-2{L^3r^3\over\beta}h_1'=0.\eqno(5.23)$$

Now we are able to calculate the metric functions. We eliminate $g_2$ in (5.22) by means of (5.14) and obtain the following linear equation for $g_1(r)$ alone
$$g_1'\B({5\over 3}Lr+{10\over 3Lr}+{4Lr\over 3\beta}\B)-{20\over 3Lr^2}g_1=L^2\B({1\over 3}-{1\over 3\beta}+{20\over L^2r^2}\B).\eqno(5.24)$$
This equation can be solved by quadratures which is considered in the appendix. Similarly we substitute (5.15) into (5.23) and get an equation for $h_2$ alone:
$$h_2'+{h_2\over r}{3L^6r^6-19L^4r^4-7L^2r^2-1\over\beta(3L^2r^2+1)^2}-{q_0L^5r^4\over (3L^2r^2+1)^2}(3L^2r^2-1)=0\eqno(5.25)$$
where
$$q_0=72\pi Gp_0.\eqno(5.26)$$
This equation, too, is solved in the appendix.

\section{Calculation of the energy-momentum tensor}

After the metric functions have been calculated the remaining two Einstein's equations determine the first order contributions to the energy-momentum tensor. We recall our results for the energy-momentum tensor of normal matter in lowest order
$$T^m_{00}=\ro_m(u_0)^2={1\over 6\pi Gt^2}+O(\al^{-1})\eqno(6.1)$$
$$T^m_{01}=\ro_m u_0u_1={1\over 6\pi Gt}{L^2r\over\beta}+O(\al^{-1})\eqno(6.2)$$
$$T^m_{11}=\ro_m(u_1)^2={1\over 6\pi G}{L^4r^2\over\beta^2}+O(\al^{-1}).\eqno(6.3)$$
This is of order $\al^0$ only, therefore, the expansion in the inhomogeneous universe is mainly driven by the gravitational field in vacuum.

To determine the observable quantities we divide (6.3) by (6.2)
$${u_1\over u_0}={L^2rt\over\beta}.\eqno(6.4)$$
The 4-velocity is normalized according to
$$g^{00}(u_0)^2+2g^{01}u_0u_1+g^{11}(u_1)^2=1=$$
$$={a^2\over D}(u_0)^2+2{b\over D}u_0u_1-{1\over D}(u_1)^2.\eqno(6.5)$$
Dividing this by $(u_0)^2$ and substituting (6.4) we get
$${1\over (u_0)^2}={a^2\over D}+2{b\over D}{L^2rt\over\beta}-{1\over D}{L^4r^2t^2\over\beta^2}=$$
$$={1\over\beta}+{2\over\al}{L^3r^2\over\beta^2}t^{1/3}+O(\al^{-2}).\eqno(6.6)$$
Multiplying by $T^m_{00}$ we obtain the matter density
$$\ro_m=\B({1\over 6\pi Gt^2}+O(\al^{-1})\B)\B({1\over\beta}+{2\over\al}{L^3r^2\over\beta^2}t^{1/3}\B)=$$
$$={1\over 6\pi Gt^2\beta}+O(\al^{-1}).\eqno(6.7)$$
This differs from the Einstein - de Sitter value by the factor $1/\beta$.

Let us also compute 4-velocity. Since we know $\ro_m$ we can calculate
$$(u_1)^2={T^m_{11}\over\ro_m}={L^4\over\beta}r^2t^2$$
or
$$u_1={L^2rt\over\sqrt{\beta}}+O(\al^{-1}).\eqno(6.8)$$
The zeroth component follows from (6.4)
$$u_0=\sqrt{\beta}+O(\al^{-1}).\eqno(6.9)$$

Now we turn to first order. We write
$$u_0=u_{00}+u_{01},\quad u_1=u_{10}+u_{11}\eqno(6.10)$$
where $u_{00}$ and $u_{10}$ are given by (6.8-9) and we use the normalization (6.5) again.
In first order $O(\al^{-1})$ we obtain
$$0=2{a_0^2\over D_0}u_{00}u_{01}+2{a_oa_1\over D_0}(u_{00})^2+O(\al^{-2})+$$
$$+\B({1\over D}\B)_1[a_0^2(u_{00})^2+2b_0u_{00}u_{10}-(u_{10})^2]\eqno(6.11)$$
because the second and third terms in (6.5) are one or two orders smaller than the first. The square bracket in (6.11) is equal to $D_0$ by zero order normalization. Therefore we arrive at
$${2\over\beta}u_{00}u_{01}+{2\over\beta a_0}a_1(u_{00})^2-{2\over D_0}(a_0a_1+b_0b_1)=0.\eqno(6.12)$$
This allows to calculate
$$u_{01}={1\over\sqrt{\beta}a_0}(a_1+Lrb_1)-\sqrt{\beta}{a_1\over a_0}.\eqno(6.13)$$

Now we turn to $G_{00}$ in first order. From the terms up to $O(\al^{-1})$ in (2.16) we find
$$(G_{00})_1={1\over D_0^2}\B[2\dot a_0\dot a_1(3a_0^2+5b_0^2)+\dot a_0^2(6a_0a_1+10b_0b_1)+$$
$$+2\dot a_1\dot b_0{b_0^3\over a_0}+2\dot b_1\dot a_0{b_0^3\over a_0}+3b_1\dot a_0\dot b_0{b_0^2\over a_0}-a_1{b_0^3\over a_0^2}\dot a_0\dot b_0\B]-$$
$$-{4\over D_0^2a_0}{a_1+Lrb_1\over\beta}\B[\dot a_0^2(3a_0^2+5b_0^2)+2\dot a_0\dot b_0{b_0^3\over a_0}\B]-4b_1{\ddot a_0b_0\over a_0D_0}+2{a_1\over a_0^2}{\ddot a_0b_0^2\over D_0}+$$
$$+4{\ddot a_0b_0^2\over a_0^2D_0}{a_1+Lrb_1\over\beta}+{2\over rD_0^2}\B(-\dot b_0b_0^2+4\dot a_0a_0b_0+6\dot a_0{b_0^3\over a_0}\B)+2{\dot a_0a_0\over D_0^2}La_0.\eqno(6.14)$$
Here we have used again $a_0'=0$ and $\ddot a_1=0$. This yields
$$(G_{00})_1={1\over a_0}\B[{g_1\over t}\B({1\over 3}+{1\over\beta}-{1\over 3\beta^2}\B)+{4\over 3}{g_2\over t}{L^3r^3\over\beta^2}+$$
$$+\B({g_1\over t}+{h_1\over t^2}\B)\B({8\over 9\beta^2}-{20\over 9\beta}-{4\over 3}\B)+Lr\B({g_2\over t}+{h_2\over t^2}\B){8\over 9}
\B({1\over\beta^2}-{1\over\beta}\B)+{20L\over 3\beta t}\B]=$$
$$=8\pi G\B[{3p_0\over a_0t^2}(1+{4\over 3}L^2r^2)+\beta\ro_{m1}+2\sqrt{\beta}{u_{01}\over 6\pi Gt^2\beta}\B].\eqno(6.15)$$
We substitute $u_{01}$ (6.13) and then the only unknown is the first order matter density $\ro_{m1}$. Collecting the many terms we obtain
$$8\pi G\beta a_0\ro_{m1}={g_1\over t}\B({5\over 3}-{35\over 9\beta}+{5\over 9\beta^2}\B)-{g_2\over t}{Lr\over 9}\B({20\over\beta}
+{4\over\beta^2}\B)+$$
$$+{h_1\over t^2}\B({4\over 3}-{44\over 9\beta}+{8\over 9\beta^2}\B)+{h_2\over t^2}{Lr\over 9}\B({8\over\beta^2}-{32\over 9\beta}\B)
+{20L\over 3\beta t}-8\pi G{3p_0\over t^2}(1+{4\over 3}L^2r^2).\eqno(6.16)$$

There remains to calculate $u_{11}$ which follows from the last Einstein's equation for $G_{01}$. We return to (5.16) and compute the first order.
We use
$$\B({1\over D}\B)_1=-{2\over D_0^2}(a_0a_1+b_0b_1)\eqno(6.17)$$
and
$$\B({1\over D^2}\B)_1=-{4\over D_0^3}(a_0a_1+b_0b_1).$$
Inserting the zero order results and collecting the terms we find
$$\B({1\over b}G_{01}+{G_{22}\over a^2r^2}\B)_1={1\over a_0^2}\B\{-{2\over Lr}\dot a'_1-{\dot b'_1\over\beta}+{a'_1\over t}\B({4\over 3}{Lr\over\beta}+{Lr\over\beta^2}+{4\over 3\beta^2Lr}\B)+$$
$$+{b'_1\over t}{2\over 3\beta}+{2L\over\beta^2}\dot a_1+{\dot b_1\over r\beta}-{a_1\over t}\B({20\over 3}{L\over\beta^2}
+{4\over 3}{L^3r^2\over\beta^2}\B)+$$
$$+{b_1\over t}\B({2\over 3r\beta^2}+{10\over 3}{L^2r\over\beta^2}-{8L\over\beta^2}\B)+L^2\B({1\over\beta^2}+{1\over\beta}\B)
\B\}=8\pi G\B({\ro_m\over b}u_0u_1\B)_1.\eqno(6.18)$$
Here the right-hand side is equal to
$$8\pi G\B[\ro_{m1}{u_{00}u_{10}\over b_0}-\ro_{m0}{b_1\over b_0^2}u_{00}u_{10}+$$
$$+{\ro_{m0}\over b_0}(u_{01}u_{10}+u_{00}u_{11})\B].\eqno(6.19)$$
Since everything except $u_{11}$ is known here we can determine it. We see that it is of order $\al^{-1}$. As a consequence all field equations are  satisfied in first order.

\section{Discussion}

We now want to put some numbers in for our present Universe. The most interesting quantity is the density of ordinary matter (6.7).
The factor $\beta=L^2r^2+1$ in (6.7) enables us to fit any value of the matter density. Let us assume a ``realistic'' density of normal matter
$$\ro_m=0.01\times\ro_{\rm crit}\eqno(7.1)$$
with a critical density
$$\ro_{\rm crit}=1.878\times 10^{-29} h^2 g/cm^3\eqno(7.2)$$
and a Hubble constant $h=0.7$ in the usual unit [6]. Taking an age $T=14\times 10^9$ years of the Universe, this corresponds to a rather small value
$$\beta=47.3\quad {\rm or}\quad LR=7.6\eqno(7.3)$$
where $R$ is the distance of the Milky Way from the origin $r=0$ where the Big Bang has taken place. To determine this distance $R$ we must use some other observable to fix the integration constant $L$. This will be done in a later paper where we work out the redshift - distance relation for the inhomogeneous universe. {\it But the small value of $LR$ seems to suggest that we live not far away from $r=0$}. The appearance of $\beta$ in (6.7) can be traced back to the non-diagonal element $b(t,r)$ in the metric and to the radial motion of the matter. If one uses comoving coordinates this motion is transformed away and then the matter density is a big problem. But we must be aware that (7.3) is not more than a lowest order orientation,
because first order may strongly change the picture.

Next we consider the radial velocity $v_m$ of the galaxies. We restrict to lowest order only. To determine $v_m$ from $u^1=u^0v_m$ we need the components with upper indices. We find
$$u^0=g^{00}u_0+g^{01}u_1={1\over\sqrt{\beta}}+{L^3r^2t\over a_0\beta^{3/2}}=$$
$$={1\over\sqrt{\beta}}+O(\al^{-1}).\eqno(7.4)$$
and
$$u^1=g^{11}u_1+g^{10}u_0={Lr\over a_0\sqrt{\beta}}-{L^2rt\over a_0^2\beta^{3/2}}=$$
$$={Lr\over\al\sqrt{\beta}}t^{-2/3}+O(\al^{-2}).\eqno(7.5)$$
At present time $t=T$, $v_m$ is of the same order of magnitude as the local velocity of the Galaxy due to gravitational attraction from nearby galaxy clusters [6] and therefore cannot be measured easily. To have simple numbers let us assume a radial velocity of 300 km/sec, so that $v_m=0.001$ because the light speed is $c=1$. Then
$$v_m={LR\over a(T)}\eqno(7.6)$$
and using (7.3) we get
$$a(T)={LR\over v_m}=7.58\times 10^3\gg 1.\eqno(7.7)$$
This value of the spatial scale function is the relevant quantity in our perturbative scheme. Since $1/a(T)\ll 1$ this scheme is consistent in the late Universe. One should remember that we have discussed the lowest order results only, we expect considerable changes in the next order.

The radiation constant $p_0$ in (5.10) is directly related to the energy density of CMB, or to the temperature $T=2.725$ K due to the Stephan-Boltzmann law. So the only constants of integration which are not known at present are $L$ and $R$ separately. Of course the redshift - distance relation will give further interesting information. This will be investigated in a later paper.

\appendix
\section{Appendix}

We first solve eq.(5.25) which is of the form
$$h'_2+f(r)h_2=g(r).\eqno(A.1)$$
It is well known that the solution of this linear equation is given by
$$h_2(r)=e^{-F}(A+\int\limits^r g(r')e^Fdr')\eqno(A.2)$$
with
$$F(r)=\int\limits^r f(r')\,dr'\eqno(A.3)$$
and $A$ is a constant of integration. According to (A.2) we must calculate the integral
$$F(r)={1\over 18}\int dx{3x^3-19x^2-7x-1\over x(x+1)(x+1/3)}$$
where we have used the substitution
$$x=L^2r^2,\quad {dr\over r}={dx\over 2x}.\eqno(A.4)$$
After decomposition into partial fractions we can integrate:
$$F(r)={1\over 18}\int\B(-
{9\over x}+{36\over x+1}-{24\over x+1/3}+{1\over (x+1/3)^2}\B)dx$$
$$=-{1\over 2}\log x+2\log (x+1)-{4\over 3}\log (x+1/3)-{1\over 18(x+1/3)}.\eqno(A.5)$$
This gives
$$e^F={(x+1)^2\over\sqrt{x}(x+1/3)^{4/3}}\exp\B(-{1\over 18(x+1/3)}\B).\eqno(A.6)$$

The remaining integral in (A.2) cannot be expressed in terms of elementary functions. Therefore we perform an expansion for $x\gg 1$.  This is not bad because we know from (7.3) that in the present Universe we have $x=L^2R^2=46.3$. One finds
$$e^F=x^{1/6}\B(1+{3\over 2x}+O(x^{-2})\B).\eqno(A.7)$$
The right side $g(r)$ in (A.1) is equal to
$$g(r)=q_0L^5r^4{(3L^2r^2-1\over (3L^2r^2+1)^2}.\eqno(A.8)$$
After expansion for large $x$ and multiplying by (A.6) we can integrate
$$\int\limits^r ge^F={q_0\over 16}x^{8/3}\B(1+{4\over 5x}+O(x^{-2})\B).\eqno(A.9)$$
This finally gives
$$h_2(r)={q_0\over 16}x^{17/6}\B(1-{7\over 10x}+O(x^{-2})\B)+H_2x^{1(6}\B(1-{3\over 2x}+O(x^{-2})\B)\eqno(A.10)$$
where $H_2$ is a constant of integration.

The equation (5.24) is solved in exactly the same way. We only give the results in the expanded form:
$$F={2\over x}\B(1-{7\over 5x}+O(x^{-2})\B)\eqno(A.11)$$
$$\int\limits^rge^F={L\over 10}\B(\log x-{291\over 5x}+O(x^{-2})\B)\eqno(A.12)$$
$$g_1(r)=\B(1-{2\over x}+O(x^{-2})\B)\B[G_1+{L\over 10}\B(\log x-{291\over 5x}+O(x^{-2})\B)\B].\eqno(A.13)$$
The leading terms in the present Universe are
$$Tg_1(R)=G_1T+{LT\over 10}\B(\log (L^2R^2)-{291\over 5L^2R^2}\B).\eqno(A.14)$$
This must be dimensionless. Since $L$ has dimension of an inverse length there is a factor $c$ (light speed) in the second term when physical units are used. To compare this first order contribution with the zeroth order $a_0$ we need the fundamental constant $L$. To get this we must investigate the redshift - distance relation, this will be done elsewhere.


\begin{thebibliography}{} 

\bibitem{} Thirring W., Classical field theory 1979, Springer, New York

\bibitem{} LUX collaboration 2013, arXiv 1310.8214, 

Science, 2015, 349, 851

\bibitem{} Scharf G., Non-standard general relativity 2012, arXiv 1208.3749.

\bibitem{} Scharf G., Non-standard cosmology 2013, arXiv 1309.5444 

\bibitem{} Kolb E.W., Matarese S., Riotto A., 2006, New J.Phys. 8, 322, 

arXiv:astr-ph/0506534

Barausse E., Matarese S., Riotto A., 2005, Phys.Rev. D 71, 063537.

arXiv:astr-ph/0501152

\bibitem{} Weinberg S. 2008, Cosmology, Oxford University Press 

\bibitem{} Weinberg S. 1972, Gravitation and Cosmology, John Wiley \& Sons 

\bibitem{} Bolejko K., Celerier M., Krasinski A.,2011, arXiv 1102.1449

Buchert T., Nayet Ch., Wiegand A. 2013, arXiv 1303.6193

\bibitem{}  Ellis G.F.R., Maartens R., MacCallum M.A.H., Relativistic Cosmology,

Cambridge University Press 2012

\bibitem{} Moseley S.H., Science 2014, 346, 696

\end{thebibliography}
\end{document}